\documentclass[]{spie}  %>>> use for US letter paper

\usepackage{amsmath,amsfonts,amssymb,mathrsfs}
\usepackage{graphicx}
\usepackage[colorlinks=true, allcolors=blue]{hyperref}
\usepackage{astro_bib_macro} %shortcuts of journals defined in this file

\title{Redundant Apodized Pupils (RAP) for high-contrast imagers robust to segmentation-due aberrations and island effects}

\author{Lucie Leboulleux\supit{a},  Alexis Carlotti\supit{a}, Mamadou N’Diaye\supit{b}, Faustine Cantalloube\supit{c}, Julien Milli\supit{a}, Arielle Bertrou-Cantou\supit{d}\supit{e}, David Mouillet\supit{a}, Nicolas Pourré\supit{a}, Christophe Vérinaud\supit{f}
\skiplinehalf
\supit{a} Univ. Grenoble Alpes, CNRS, IPAG, 38000 Grenoble, France,\\
\supit{b} Université Côte d’Azur, Observatoire de la Côte d’Azur, CNRS, Laboratoire Lagrange, France,\\
\supit{c} Aix Marseille Université, CNRS, LAM (Laboratoire d’Astrophysique de Marseille) UMR 7326, 13388 Marseille, France,\\
\supit{d} LESIA, Observatoire de Paris, Université PSL, CNRS, Sorbonne Université, Université de Paris, 5 place Jules Janssen, 92195 Meudon, France,\\
\supit{e} Department of Astronomy, California Institute of Technology, Pasadena, CA 91125, USA,\\
\supit{f} European Southern Observatory (ESO), Karl-Schwarzschild-Str. 2, D-85748 Garching bei Muenchen, Germany
}

\authorinfo{Further author information, send correspondence to Lucie Leboulleux: E-mail: lucie.leboulleux@univ-grenoble-alpes.fr}

\begin{document} 
\maketitle

\begin{abstract}

The imaging and characterization of a larger range of exoplanets, down to young Jupiters and exo-Earths will require accessing very high contrasts at small angular separations with an increased robustness to aberrations, three constraints that drive current instrumentation development.

This goal relies on efficient coronagraphs set up on extremely large diameter telescopes such as the Thirty Meter Telescope (TMT), the Giant Magellan Telescope (GMT), or the Extremely Large Telescope (ELT). However, they tend to be subject to specific aberrations that drastically deteriorate the coronagraph performance: their primary mirror segmentation implies phasing errors or even missing segments, and the size of the telescope imposes large spiders, generating low-wind effect as already observed on the Very Large Telescope (VLT)/SPHERE instrument or at the Subaru telescope, or adaptive-optics-due petaling, studied in simulations in the ELT case. The ongoing development of coronagraphs has then to take into account their sensitivity to such errors.

We propose an innovative method to generate coronagraphs robust to primary mirror phasing errors and low-wind and adaptive-optics-due petaling effect. This method is based on the apodization of the segment or petal instead of the entire pupil, this apodization being then repeated to mimic the pupil redundancy.

We validate this so-called Redundant Apodized Pupil (RAP) method on a James Webb Space Telescope-like pupil composed of 18 hexagonal segments segments to align, and on the VLT architecture in the case of residual low-wind effect.

\end{abstract}


%%%%%%%%%%%%%%%%%%%%%%%%%%%%%%%%%%%%%%%%%%%%%%%%%%%%%%%%%%%%%%%%%
%%%%%%%%%%%%%%%%%%%%%%%%%%%%%%%%%%%%%%%%%%%%%%%%%%%%%%%%%%%%%%%%%
%%%%%%%%%%%%%%%%%%%%%%%%%%%%%%%%%%%%%%%%%%%%%%%%%%%%%%%%%%%%%%%%%
\section{INTRODUCTION}
\label{sec:INTRODUCTION}
%%%%%%%%%%%%%%%%%%%%%%%%%%%%%%%%%%%%%%%%%%%%%%%%%%%%%%%%%%%%%%%%%
%%%%%%%%%%%%%%%%%%%%%%%%%%%%%%%%%%%%%%%%%%%%%%%%%%%%%%%%%%%%%%%%%
%%%%%%%%%%%%%%%%%%%%%%%%%%%%%%%%%%%%%%%%%%%%%%%%%%%%%%%%%%%%%%%%%

The next challenges in high-contrast imaging rely on the development of imagers set up on giant primary mirror telescopes: the Extremely Large Telescope (ELT, 39 m), the Thirty Meter Telescope (TMT, 30 m), the Giant Magellan Telescope (GMT, 24.5 m), the space Large UV-Optical-InfraRed telescope (LUVOIR, 6 m)... Due to their large diameters, these telescopes will include a segmented primary mirror, from 7 circular segments (GMT) to 798 hexagonal ones (ELT). However, this segmentation will be subject to segment misalignements and instabilities that have to be precisely controlled to reach the contrasts deep enought to access their scientific targets \cite{Yaitskova2002, Yaitskova2003, Stahl2017, Stahl2020, Leboulleux2018, Leboulleux2018a, Laginja2021}.

In addition, the secondary mirrors of the ELT and TMT will be maintained by large spiders, that will also be responsible for aberrations called island effects. First, the low-wind effect, already detected on the Spectro Polarimetric High contrast Exoplanet REsearch (SPHERE) instrument at the Very Large Telescope (VLT) and at the Subaru Telescope, is due to thermal gradients along the spiders for windspeed lower than 1 m/s and corresponds to local pistons, tips, and tilts on the petals delimited by the spiders \cite{Sauvage2016, Milli2018}. Secondly, the petaling, only predicted by numerical simulations so far, happens when the spiders are larger than the atmospheric turbulence Fried parameter $r_0$ and disturb the wavefront error reconstruction by the adaptive optics system: this generates piston discontinuities between petals \cite{Bertrou-Cantou2020, Bertrou-Cantou2022}.

These aberrations have low-order components with impacts close to the optical axis. Therefore, coronagraphs are highly sensitive to them: on LUVOIR, maintaining a $10^{-10}$ constrast requires a segment phasing down to $10$ pm RMS \cite{Laginja2021, TheLUVOIRTeam2019}, and on VLT/SPHERE, the low-wind effect generated a phase error of $200$ nm RMS before the 2017 coating, degrading the contrast by a factor of about $50$ at $0.1$'' \cite{Sauvage2015, Sauvage2016, Cantalloube2019}.

In this context, the robustness of the coronagraph to such errors is critical and is evaluated during the design process before validating a final design. We have developed a procedure to directly integrate this robustness criterion in the coronagraph design process and generate specific coronagraphs robust to segmentation-due errors and island effects. This method derives from the Pair-based (PASTIS), an analytical expression of the coronagraphic Point Spread Function (PSF) behind a segmented mirror \cite{Leboulleux2018, Leboulleux2018a, Laginja2021} and relies on the development of apodizers at the scale of the segment or petal instead of the overall pupil. These apodizers are then repeated to mimic the pupil geometry and generate the so-called Redundant Apodized Pupils (RAP) \cite{Leboulleux2022, Leboulleux2022a}. In this proceeding, we apply this RAP methodology to other pupils and specifications. The section \ref{sec:CONCEPT OF REDUNDANT APODIZED PUPILS} of this paper intends to introduce the RAP concept and methodology and sections \ref{sec:APPLICATION TO SEGMENTATION-DUE ERRORS} and \ref{sec:APPLICATION TO ISLAND EFFECTS} to validate it respectively for segmentation-induced errors (segment phasing) on a James Webb Space Telescope (JWST)-like pupil and for island effects (in particular low-wind effect) on the VLT pupil.

%%%%%%%%%%%%%%%%%%%%%%%%%%%%%%%%%%%%%%%%%%%%%%%%%%%%%%%%%%%%%%%%%
%%%%%%%%%%%%%%%%%%%%%%%%%%%%%%%%%%%%%%%%%%%%%%%%%%%%%%%%%%%%%%%%%
%%%%%%%%%%%%%%%%%%%%%%%%%%%%%%%%%%%%%%%%%%%%%%%%%%%%%%%%%%%%%%%%%
\section{CONCEPT OF REDUNDANT APODIZED PUPILS}
\label{sec:CONCEPT OF REDUNDANT APODIZED PUPILS}
%%%%%%%%%%%%%%%%%%%%%%%%%%%%%%%%%%%%%%%%%%%%%%%%%%%%%%%%%%%%%%%%%
%%%%%%%%%%%%%%%%%%%%%%%%%%%%%%%%%%%%%%%%%%%%%%%%%%%%%%%%%%%%%%%%%
%%%%%%%%%%%%%%%%%%%%%%%%%%%%%%%%%%%%%%%%%%%%%%%%%%%%%%%%%%%%%%%%%

%%%%%%%%%%%%%%%%%%%%%%%%%%%%%%%%%%%%%%%%%%%%%%%%%%%%%%%%%%%%%%%%%
\subsection{Reminders}
\label{sec:Reminders}
%%%%%%%%%%%%%%%%%%%%%%%%%%%%%%%%%%%%%%%%%%%%%%%%%%%%%%%%%%%%%%%%%

The PASTIS analytical model expresses the intensity in the dark region of a coronagraph as a function of the segment phase errors, which can be any segment-level Zernike polynomial \cite{Leboulleux2017, Leboulleux2018, Laginja2021, Laginja2022} or combination of Zernike polynomials \cite{Leboulleux2018a}. In particular, for segment-level piston errors, it stipulates that the intensity in the dark zone $I$ in monochromatic light can be expressed as follows: 
\begin{equation}
\label{eq:PASTIS}
    I(\mathbf{u}) = \left \Vert \widehat{S}(\mathbf{u}) \right \Vert ^2 \times \sum_{i=1}^{n_{seg}} \sum_{j=1}^{n_{seg}} c_i a_i c_j a_j \cos((\mathbf{r_j} - \mathbf{r_i} ). \mathbf{u})
\end{equation}
$n_{seg}$ being the number of segments in the primary mirror, $\mathbf{u}$ the position vector in the focal plane, $\widehat{S}$ the Fourier Transform of the segment shape $S$, $(c_k)_{k\in \lbrack 1,n_{seg} \rbrack}$ calibration coefficients taking into account the coronagraph, $(a_k)_{k\in \lbrack 1,n_{seg} \rbrack}$ the segment-level piston coefficients, and $(\mathbf{r_k})_{k\in \lbrack 1,n_{seg} \rbrack}$ the position vectors of the segment centers. This expression corresponds to the product between a low-order envelope $\left \Vert \widehat{S}(\mathbf{u}) \right \Vert ^2$, i.e., the segment PSF, and a sum of interference fringes between all pairs of segments, modulated by the piston error coefficients.

The PASTIS procedure, including all codes, is freely available on GitHub \cite{pastis}.

%%%%%%%%%%%%%%%%%%%%%%%%%%%%%%%%%%%%%%%%%%%%%%%%%%%%%%%%%%%%%%%%%
\subsection{Principle}
\label{sec:Principle}
%%%%%%%%%%%%%%%%%%%%%%%%%%%%%%%%%%%%%%%%%%%%%%%%%%%%%%%%%%%%%%%%%

From the equation \ref{eq:PASTIS}, one can derive that modifying the low-order envelope $\left \Vert \widehat{S}(\mathbf{u}) \right \Vert ^2$ impacts the intensity in the dark region. In particular, deepening the contrast of this envelope releases the constraints on the segment-level piston coefficients $(a_k)_{k\in \lbrack 1,n_{seg} \rbrack}$. This is the basis of Redundant Apodized Pupils (RAP): to design coronagraphs robust to segment-level pistons, we propose a 2 step approach:

\textbullet~developing apodizers at the segment scale with a low-order envelope optimized in the target dark region,

\textbullet~multiplying these segment-level apodizers to reproduce the pupil segmentation pattern.

If the full pupil PSF has an angular resolution of $\lambda/D$ ($\lambda$ is the wavelength and $D$ the pupil diameter), then the low-order envelope has a resolution element of $\lambda/d$, where is the segment diameter ($d=D/N$), $N$ is the number of segments along the primary mirror diameter ($N \sim 3$ for the GMT, $\sim 5$ for the JWST...). This implies that a dark zone Inner Working Angle (IWA) (respectively Outer Working Angle or OWA) of $\alpha_I \lambda/D$ (resp. $\alpha_O \lambda/D$) corresponds to an angular separation of $(\alpha_I/N) \times \lambda/d$ (resp. $(\alpha_O/N) \times \lambda/d$) in the low-order envelope. This will be the main constraint to the RAP concept: since the low-order envelope has a resolution element of $\lambda/d = N \lambda/D$, optimizing the segment can only impact the PSF at separations higher than $N \times \lambda/D$, ie., outside the segment diffraction limit, limiting the access to small angular separations. 

More detailed explanations can be found in Leboulleux et al. 2022a,b\cite{Leboulleux2022, Leboulleux2022a} and all AMPL codes to design segment-level phase and amplitude apodizers, some output designs (fits files) and the Python codes to test them are accessible for free on PerSciDo: \url{https://perscido.univ-grenoble-alpes.fr/datasets/DS363} and \url{https://perscido.univ-grenoble-alpes.fr/datasets/DS370}.

%%%%%%%%%%%%%%%%%%%%%%%%%%%%%%%%%%%%%%%%%%%%%%%%%%%%%%%%%%%%%%%%%
%%%%%%%%%%%%%%%%%%%%%%%%%%%%%%%%%%%%%%%%%%%%%%%%%%%%%%%%%%%%%%%%%
%%%%%%%%%%%%%%%%%%%%%%%%%%%%%%%%%%%%%%%%%%%%%%%%%%%%%%%%%%%%%%%%%
\section{APPLICATION TO SEGMENTATION-DUE ERRORS}
\label{sec:APPLICATION TO SEGMENTATION-DUE ERRORS}
%%%%%%%%%%%%%%%%%%%%%%%%%%%%%%%%%%%%%%%%%%%%%%%%%%%%%%%%%%%%%%%%%
%%%%%%%%%%%%%%%%%%%%%%%%%%%%%%%%%%%%%%%%%%%%%%%%%%%%%%%%%%%%%%%%%
%%%%%%%%%%%%%%%%%%%%%%%%%%%%%%%%%%%%%%%%%%%%%%%%%%%%%%%%%%%%%%%%%

%%%%%%%%%%%%%%%%%%%%%%%%%%%%%%%%%%%%%%%%%%%%%%%%%%%%%%%%%%%%%%%%%
\subsection{Specifications and design}
\label{sec:Specifications}
%%%%%%%%%%%%%%%%%%%%%%%%%%%%%%%%%%%%%%%%%%%%%%%%%%%%%%%%%%%%%%%%%

This application case has been studied in Leboulleux et al. 2022\cite{Leboulleux2022} for a GMT-like aperture, combined with an Apodized Pupil Lyot Coronagraph (APLC) and with an Apodizing Phase Plate (APP) to generate circular symmetrical dark zones. In this section, we focus on a JWST-like pupil made of $18$ hexagonal segments ($\sim 5$ across the pupil diameter), combined with a redundant amplitude apodizer or Shaped Pupil (SP) to generate a small circular dark zone exactly at the planet location (here at $8.3 \lambda/D$). Practically, the objective would be to recover the planet spectrum with an Integral Field Spectrograph (IFS).

\begin{figure}%[h]
   \begin{center}
   \begin{tabular}{c}
   \includegraphics[width=10cm]{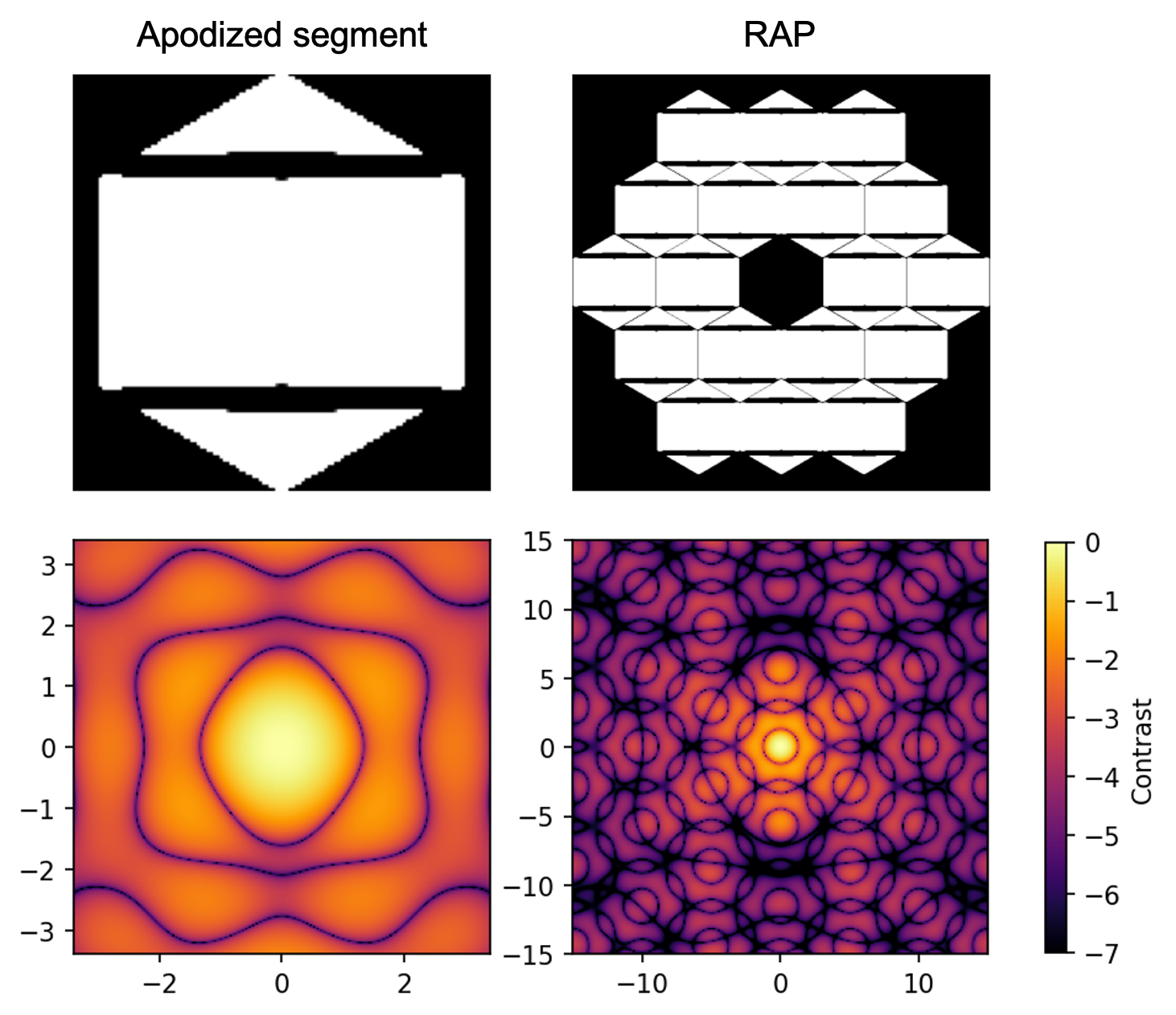}
   \end{tabular}
   \end{center}
   \caption[JWST_Fig1] 
   { \label{fig:JWST_Fig1} Shaped Pupil design from redundant apodized petals: (top) apodized petals (North and East) and recombined shaped pupil, (bottom) contrast maps in logarithmic scale generated by the apodizations of the first line.}
   \end{figure}

A SP is designed at the scale of the segment to create a small dark zone, $1/4.4 \times \lambda/d$-large, at $1.88 \lambda/d$ (see Fig. \ref{fig:JWST_Fig1}, left). The $4.4$ value is the number of segments per diameter and is more precise than $5$: it corresponds to the $(\lambda/d)/(\lambda/D)$ ratio, $\lambda/d$ being the segment spatial resolution and $\lambda/D$ the full pupil one. Once multiplied over the pupil with respect to its segmentation, this SP generates a circular $1 \lambda/D$-large dark zone, at $8.3 \lambda/D$ and down to an average contrast in this small dark zone of $9.9 \times 10^{-7}$ (see Fig. \ref{fig:JWST_Fig1}, right). This final RAP has a transmission of $87\%$ and a planetary throughput of $76\%$.

%-------------
%-------------

%%%%%%%%%%%%%%%%%%%%%%%%%%%%%%%%%%%%%%%%%%%%%%%%%%%%%%%%%%%%%%%%%
\subsection{Error budget}
\label{sec:Error budget}
%%%%%%%%%%%%%%%%%%%%%%%%%%%%%%%%%%%%%%%%%%%%%%%%%%%%%%%%%%%%%%%%%

\begin{figure}%[h]
   \begin{center}
   \begin{tabular}{c}
   \includegraphics[width=14cm]{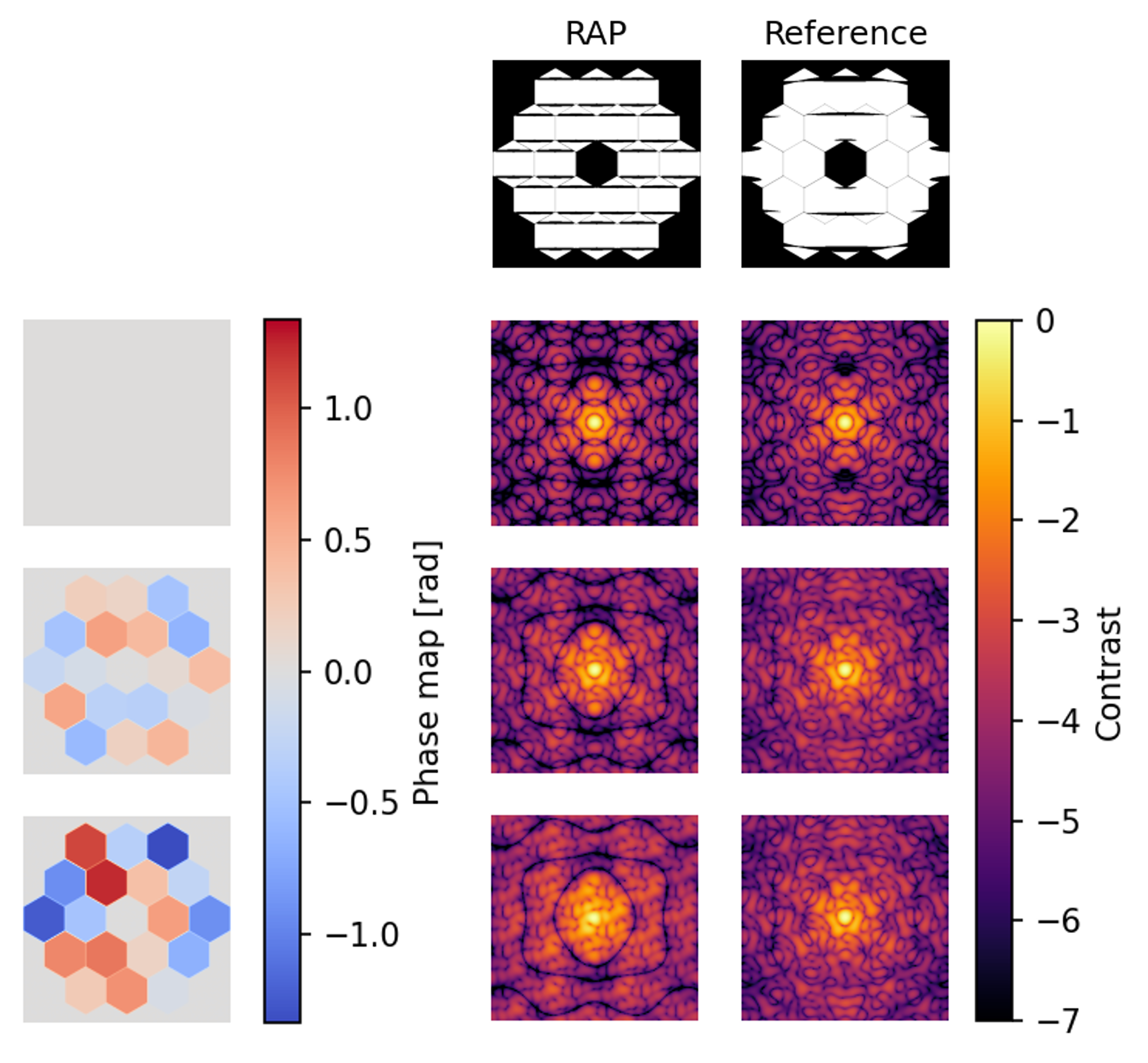}
   \end{tabular}
   \end{center}
   \caption[JWST_Fig2] 
   { \label{fig:JWST_Fig2} Impact of petal-level piston phasing errors on the coronagraphic PSF and contrast: (left) phase maps applied on the entrance pupil, from $0$ (top) to $100$ nm RMS (center) and to $200$ nm RMS (bottom), (center) coronagraphic PSFs behind the RAP in presence of the left column errors, (right) coronagraphic PSFs behind the reference SP in presence of the left column errors. The wavelength $\lambda$ is $1.6 \mu$m.}
   \end{figure}

The stability of this RAP to segment-level piston errors is compared to the one of a later-on called "reference" SP, more classically obtained by apodizing the full pupil instead of one segment only. This apodizer is visible on Fig. \ref{fig:JWST_Fig2} (top right image), has a transmission of $94\%$, a planetary throughput of $89\%$, and an average contrast in the same dark zone than the RAP one of $8.6\times 10^{-7}$.

Fig. \ref{fig:JWST_Fig2} compares the evolution of the coronagraphic PSFs and of their contrast maps when the phasing aberrations increase (no aberration, $100$ nm RMS, and $200$ nm RMS). In the RAP case, the average contrast in the small dark zone evolves from $9.9 \times 10^{-7}$ to $1.1 \times 10^{-5}$ (factor of $12$ compared to the no aberration case), while in the reference case, it evolves from $8.6 \times 10^{-7}$ to $1.8 \times 10^{-4}$ (factor of $209$ compared to the no aberration case).

%-------------
%-------------

\begin{figure}%[h]
   \begin{center}
   \begin{tabular}{c}
   \includegraphics[width=12cm]{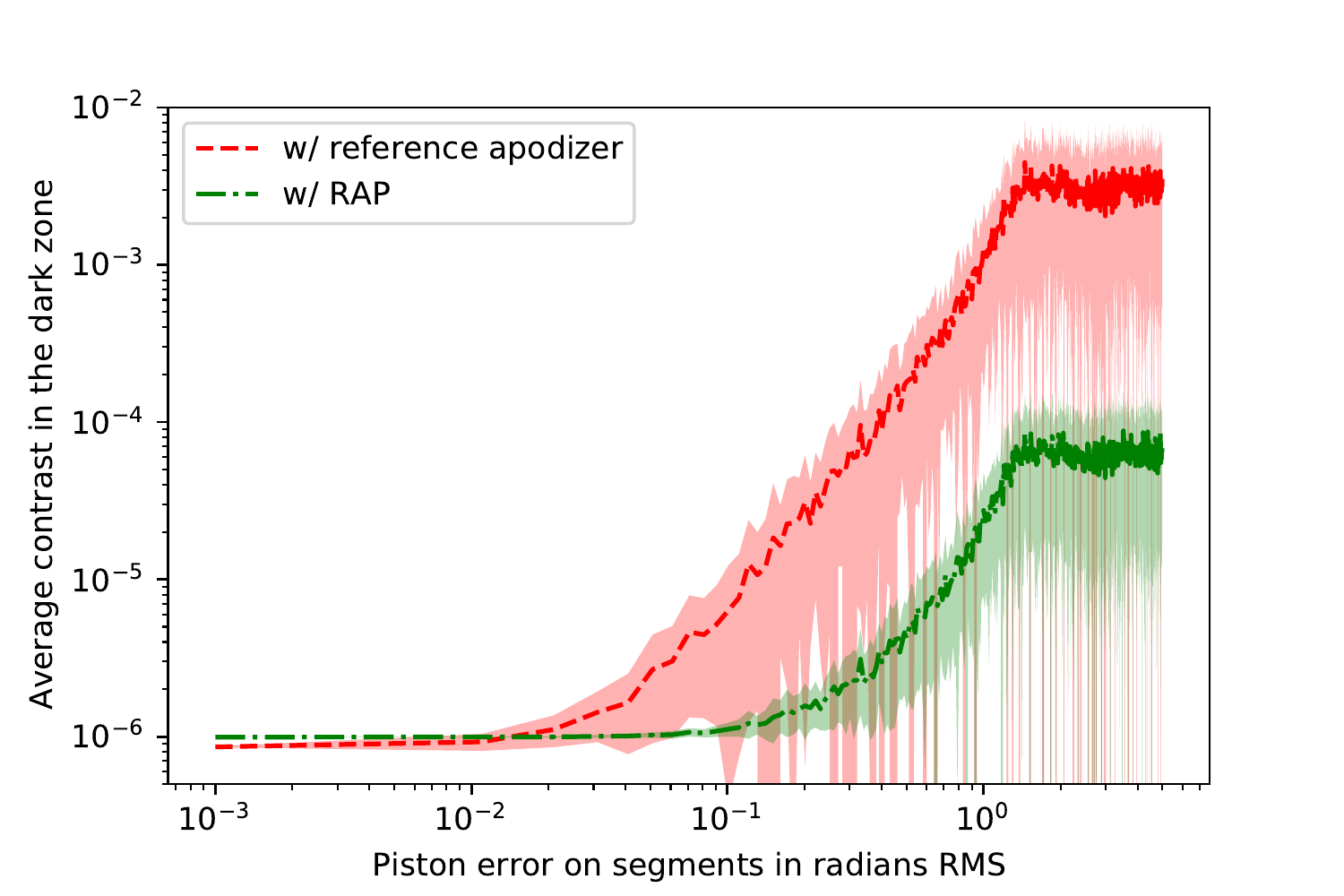}
   \end{tabular}
   \end{center}
   \caption[JWST_Fig3] 
   { \label{fig:JWST_Fig3} Evolution of the average contrast in the dark zone with the amplitude of the petal-level piston errors: for each amplitude, $50$ random phase maps are propagated through the optical system and the mean value of the $50$ resulting dark zone contrasts is computed (dark lines). The light color areas are delimited by the standard deviations of the $50$ dark zone contrasts. The red plots correspond to the reference SP case and the green ones to the RAP case.}
   \end{figure}

More generally, Fig. \ref{fig:JWST_Fig3} shows the evolution of the average contrast in the dark zone for piston-phasing errors from $1$ mrad RMS to $5$ rad RMS. From a few $0.1$ rad RMS, the RAP is at least $20$ times more robust than the reference SP: the segment low-order envelope remains in the coronagraphic PSF despite the aberrations and maintains the image intensity below $10^{-6}$ and the deterioration of contrast mostly comes from the Strehl loss of the PSF core. In the reference case, the segment apodization is not optimized to constraint this low-order envelope and the SP is then sensitive to segment-level aberrations.

%-------------
%-------------

%%%%%%%%%%%%%%%%%%%%%%%%%%%%%%%%%%%%%%%%%%%%%%%%%%%%%%%%%%%%%%%%%
%%%%%%%%%%%%%%%%%%%%%%%%%%%%%%%%%%%%%%%%%%%%%%%%%%%%%%%%%%%%%%%%%
%%%%%%%%%%%%%%%%%%%%%%%%%%%%%%%%%%%%%%%%%%%%%%%%%%%%%%%%%%%%%%%%%
\section{APPLICATION TO ISLAND EFFECTS}
\label{sec:APPLICATION TO ISLAND EFFECTS}
%%%%%%%%%%%%%%%%%%%%%%%%%%%%%%%%%%%%%%%%%%%%%%%%%%%%%%%%%%%%%%%%%
%%%%%%%%%%%%%%%%%%%%%%%%%%%%%%%%%%%%%%%%%%%%%%%%%%%%%%%%%%%%%%%%%
%%%%%%%%%%%%%%%%%%%%%%%%%%%%%%%%%%%%%%%%%%%%%%%%%%%%%%%%%%%%%%%%%

%%%%%%%%%%%%%%%%%%%%%%%%%%%%%%%%%%%%%%%%%%%%%%%%%%%%%%%%%%%%%%%%%

As a reminder, in this proceeding island effects include two categories of aberrations, depending on their cause:

- the low-wind effect: the phase is fragmented into differential piston, tip, and tilt on the pupil petals delimited by the spiders. This effect has been detected on VLT/SPHERE \cite{Sauvage2015, Sauvage2016, Lamb2017, Milli2017, Milli2018, Cantalloube2019} and at the Subaru Telescope \cite{Vievard2019} where it limits the instrument performance in low-wind conditions. It is also expected on upcoming giant telescopes like the ELT and the TMT \cite{Holzlohner2021}, two application cases of the RAP concept that are the purpose of Leboulleux et al. 2022b\cite{Leboulleux2022a}.

- the post-adaptive optics petaling effect: the phase is fragmented into differential pistons on the pupil petals. This effect has only been studied in simulations for now but is expected at the ELT \cite{Bertrou-Cantou2022}. The application of the RAP concept to the ELT Leboulleux et al. 2022b\cite{Leboulleux2022a} is also efficient to mitigate petaling effect on the ELT.

\begin{figure}%[h]
   \begin{center}
   \begin{tabular}{c}
   \includegraphics[width=12cm]{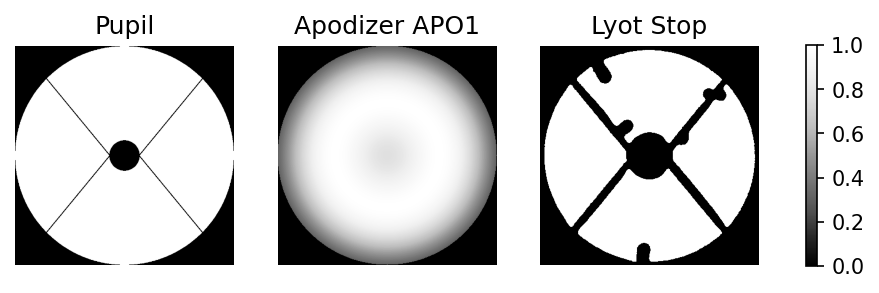}
   \end{tabular}
   \end{center}
   \caption[VLT_PupilPlanes] 
   { \label{fig:VLT_PupilPlanes} Pupil planes of the SPHERE coronagraph chosen for this study: (left) the VLT pupil, (center) the apodizer APO1 which consists in a prolate function, and (right) the Lyot Stop.}
   \end{figure}

Because of the structure of these aberrations (differential pistons on large areas of the pupil), they can be mitigated with a RAP design. In this section, we then apply it on the VLT architecture, composed of two different types of petals (North and South being identical albeit symmetrical to each other, East and West too) to mitigate low-wind effect. We compare its contrast stability with the one issued from the SPHERE instrument composed of the apodizer APO1, the 185 mas diameter focal plane mask, and the Lyot Stop of Fig. \ref{fig:VLT_PupilPlanes}.

%-------------
%-------------

\subsection{Specifications and design}
\label{sec:Specifications2}
%%%%%%%%%%%%%%%%%%%%%%%%%%%%%%%%%%%%%%%%%%%%%%%%%%%%%%%%%%%%%%%%%

The RAP design aims to dig a circular symmetrical dark zone between $8$ and $20 \lambda/D$ with a contrast of $\sim 10^{-6}$. Once again, we focus on a binary amplitude pupil mask, with no FPM and no Lyot stop (SP). 

\begin{figure}%[h]
   \begin{center}
   \begin{tabular}{c}
   \includegraphics[width=12cm]{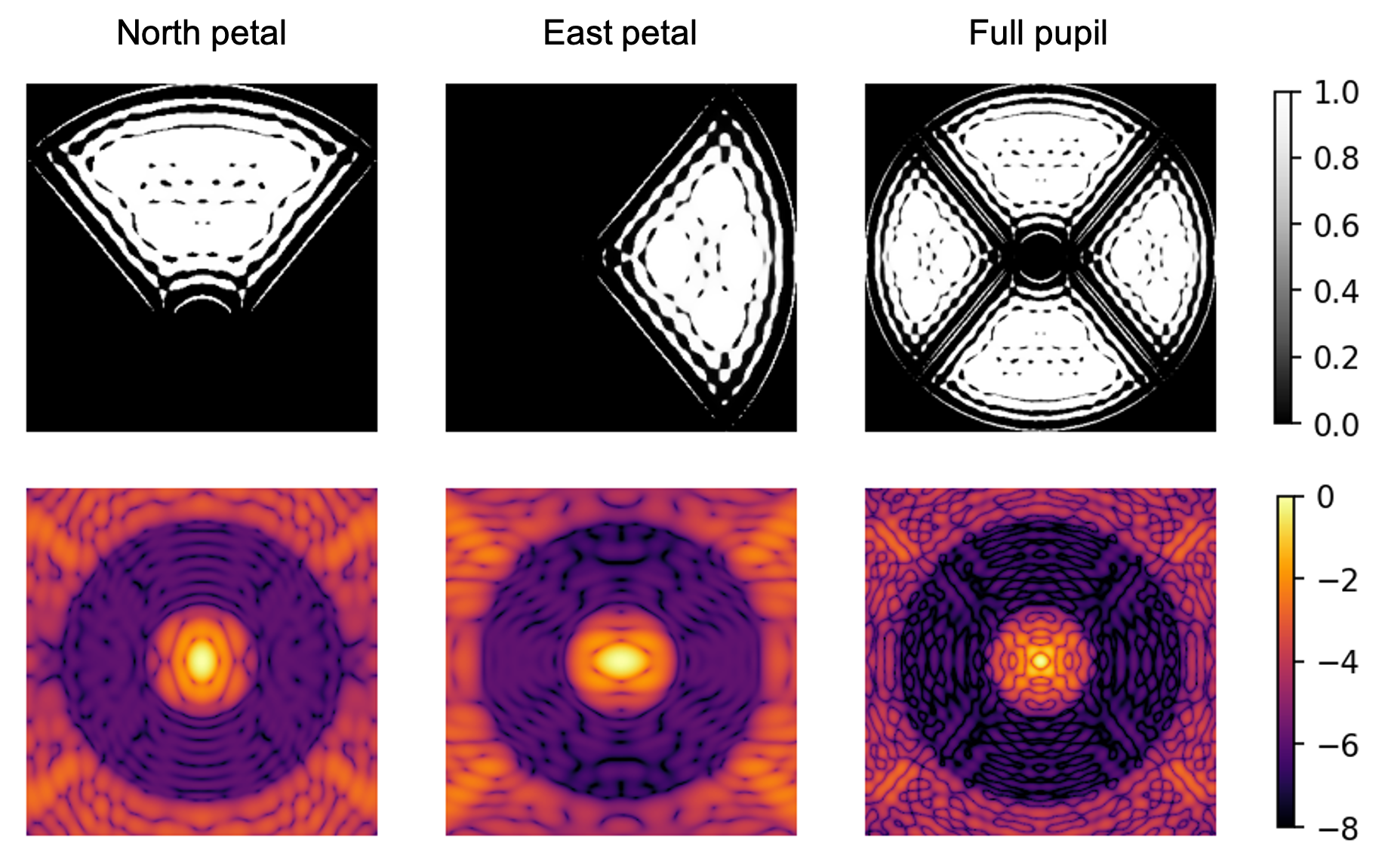}
   \end{tabular}
   \end{center}
   \caption[VLT_Fig1] 
   { \label{fig:VLT_Fig1} SP design from redundant apodized petals: (top) apodized petals (North and East) and recombined shaped pupil, (bottom) contrast maps in logarithmic scale generated by the apodizers of the first line.}
   \end{figure}

The output designs are visible in Fig. \ref{fig:VLT_Fig1}with both North and East petals and the low-order envelopes they create. The South and West petals are symmetrical to the North and East ones respectively. When combined together, the four apodized petals form the RAP, whose performance respects the specifications (contrast of $\sim 10^{-6}$ between $8$ and $20 \lambda/D$): the average contrast in the dark zone is $2.5 \times 10^{-7}$ and the maximum contrast is $1.6 \times 10^{-6}$. In addition, this RAP has a planetary throughput of $36\%$ and a transmission of $60\%$.

%-------------
%-------------

%%%%%%%%%%%%%%%%%%%%%%%%%%%%%%%%%%%%%%%%%%%%%%%%%%%%%%%%%%%%%%%%%
\subsection{Error budget}
\label{sec:Error budget2}
%%%%%%%%%%%%%%%%%%%%%%%%%%%%%%%%%%%%%%%%%%%%%%%%%%%%%%%%%%%%%%%%%

\begin{figure}%[h]
   \begin{center}
   \begin{tabular}{c}
   \includegraphics[width=12cm]{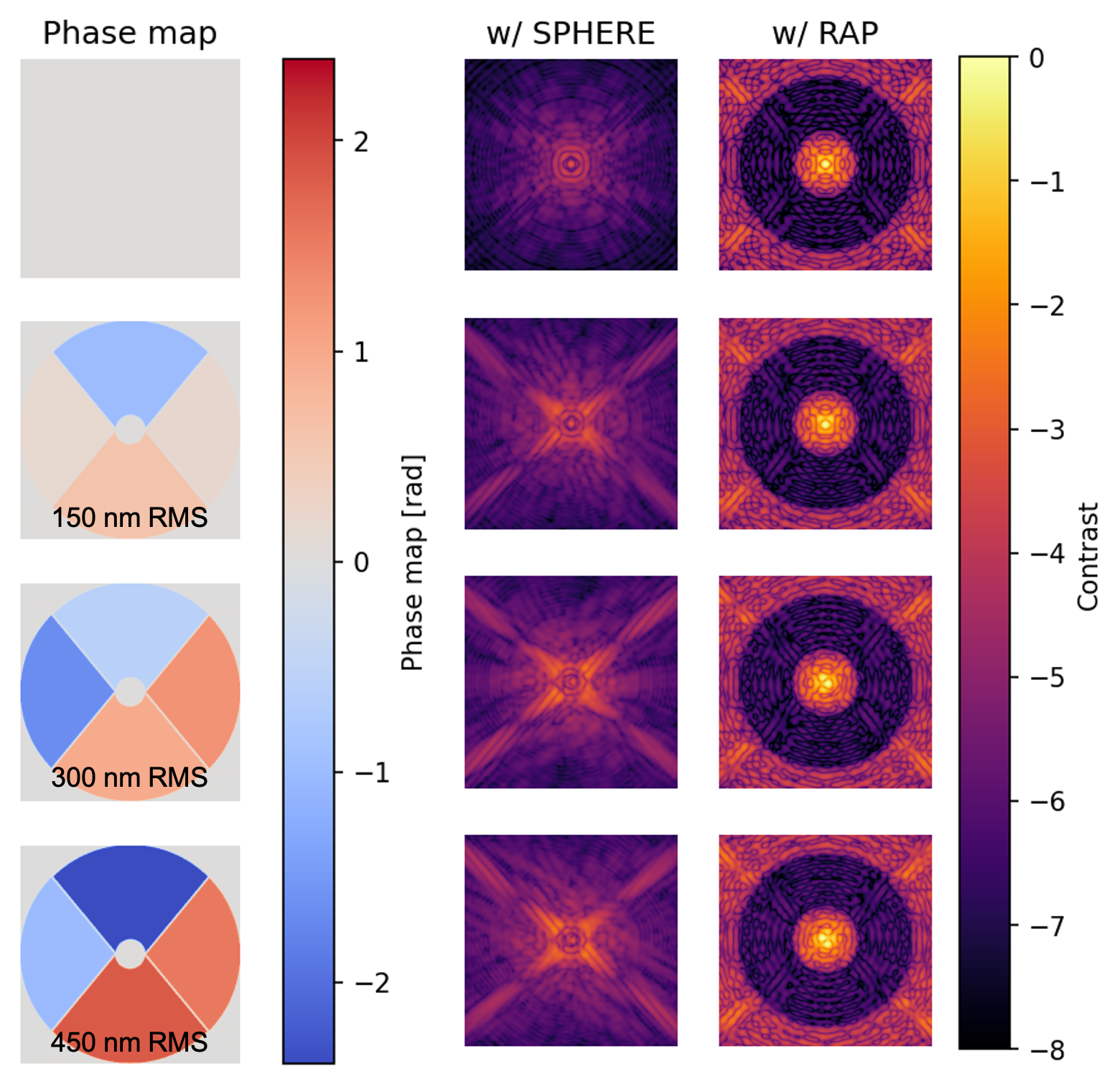}
   \end{tabular}
   \end{center}
   \caption[VLT_Fig2] 
   { \label{fig:VLT_Fig2} Impact of petal-level piston phasing errors on the coronagraphic PSF and contrast: (left) phase maps applied on the entrance pupil, from $0$ (top) to $450$ nm RMS (bottom), (center) coronagraphic PSFs behind the SPHERE coronagraph in presence of the left column errors, (right) coronagraphic PSFs behind the RAP in presence of the left column errors. The wavelength $\lambda$ is $1.6 \mu$m.}
   \end{figure}

For both the SPHERE coronagraph and the RAP, Fig. \ref{fig:VLT_Fig2} illustrates the impact of petal-level piston errors on the coronagraphic PSF and contrast, for random differential pistons of $0$, $150$, $300$, and $450$ nm RMS. The middle (respectively right) column indicates the contrast maps in the focal plane behind the SPHERE coronagraph (respectively RAP) in presence of the left column errors. In the same dark zone ($8-20 \lambda/D$), the contrast is deteriorated from $1.2 \times 10^{-6}$ to $1.2 \times 10^{-5}$ with the SPHERE coronagraph (factor of $10$), while from $2.5 \times 10^{-7}$ to $4.0 \times 10^{-7}$ with the RAP (factor of almost $2$). Despite increasing aberration amplitudes, no diffraction spikes appear in the RAP coronagraphic PSF and the deterioration of contrast is only induced by the loss of coherence of the PSF core. 

%-------------
%-------------

\begin{figure}%[h]
   \begin{center}
   \begin{tabular}{c}
   \includegraphics[width=12cm]{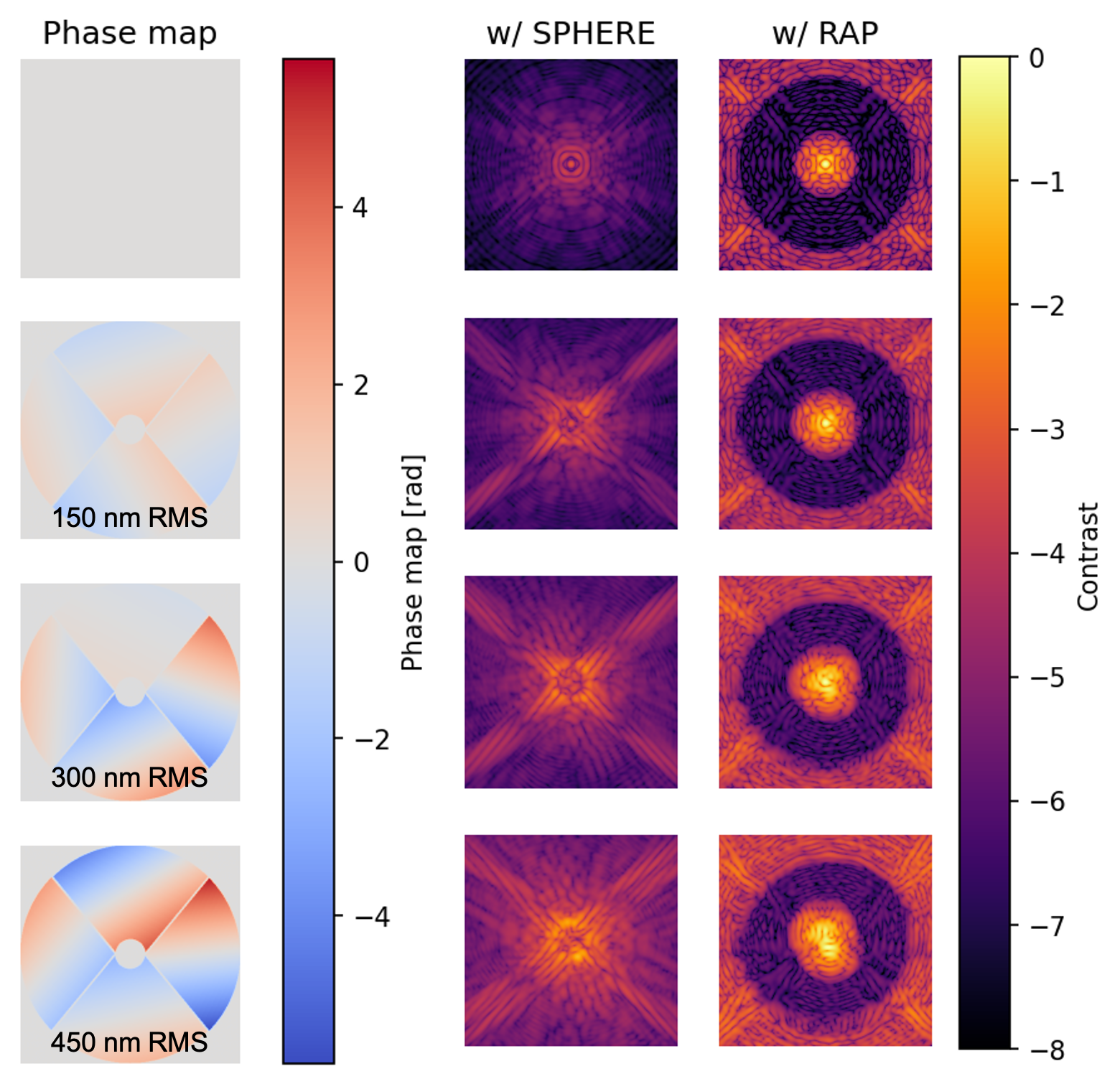}
   \end{tabular}
   \end{center}
   \caption[VLT_Fig3] 
   { \label{fig:VLT_Fig3} Impact of petal-level tip-tilt phasing errors on the coronagraphic PSF and contrast: (left) phase maps applied on the entrance pupil, from $0$ (top) to $450$ nm RMS (bottom), (center) coronagraphic PSFs behind the SPHERE coronagraph in presence of the left column errors, (right) coronagraphic PSFs behind the RAP in presence of the left column errors. The wavelength $\lambda$ is $1.6 \mu$m.}
   \end{figure}

We now test the robustness of the RAP apodizer to tip-tilt petal-level errors, which are also induced by the low-wind effect. Fig. \ref{fig:VLT_Fig3} illustrates the impact of such errors, from $0$ to $450$ nm RMS, on the coronagraphic PSFs behind both the RAP and the SPHERE coronagraph. In the same dark zone ($8-20 \lambda/D$), the contrast is deteriorated from $1.2 \times 10^{-6}$ to $5.9 \times 10^{-5}$ with SPHERE (factor of almost $50$), while from $2.5 \times 10^{-7}$ to $1.2 \times 10^{-5}$ with the RAP (factor of almost $50$), with a high impact at small angular separations. However, these degradation factors and the morphology of the coronagraphic PSFs evolve a lot from one phase map case to the other, indicating an equivalent contrast robustness between both designs.

%-------------
%-------------

\begin{figure}%[h]
   \begin{center}
   \begin{tabular}{c}
   \includegraphics[width=12cm]{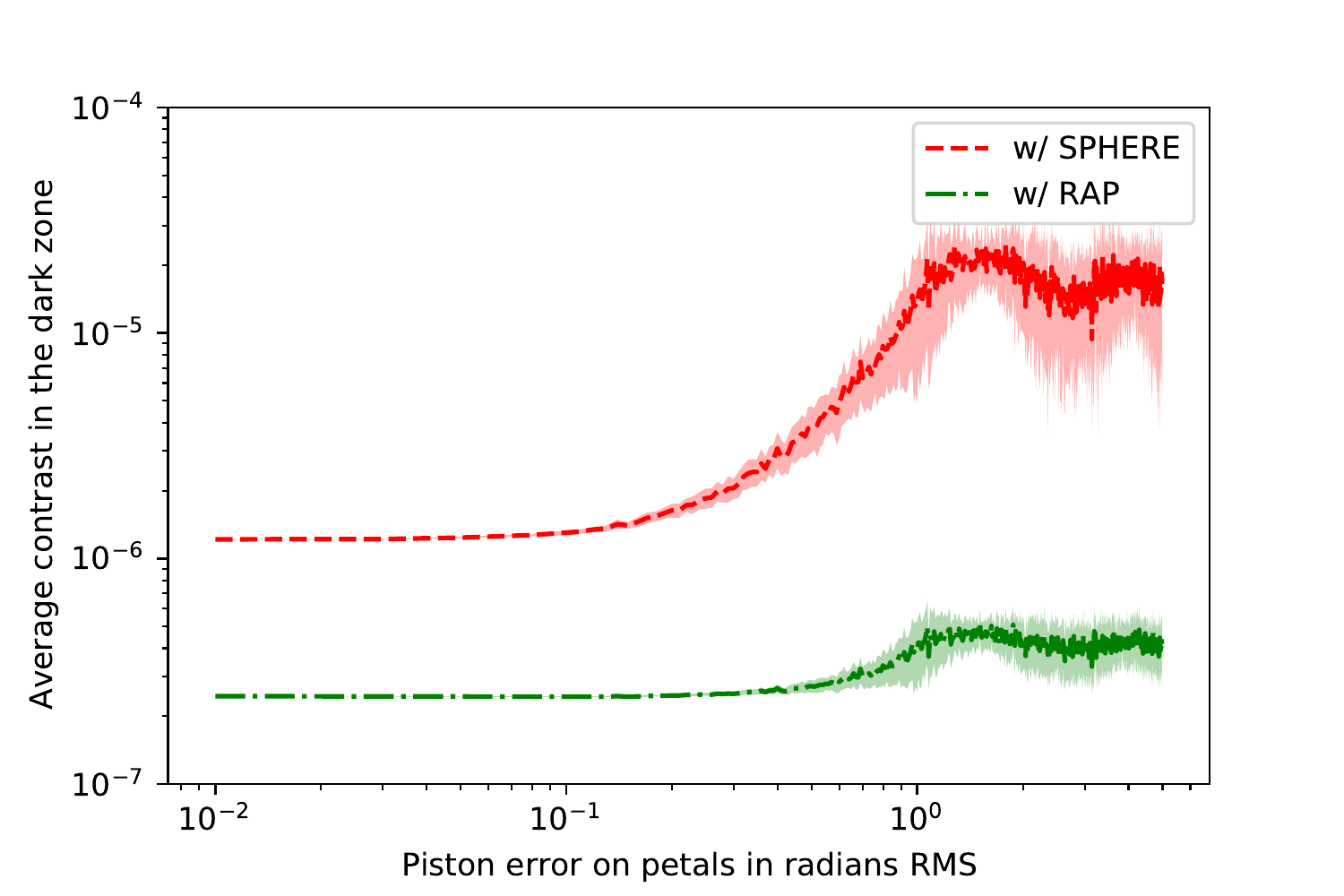}
   \end{tabular}
   \end{center}
   \caption[VLT_Fig4] 
   { \label{fig:VLT_Fig4} Evolution of the average contrast in the dark zone with the amplitude of the petal-level piston errors: for each amplitude, $25$ random phase maps are propagated through the optical system and the mean value of the $25$ resulting dark zone contrasts is computed (dark lines). The light color areas are delimited by the standard deviations of the $25$ dark zone contrasts. The red plots correspond to the SPHERE case and the green ones to the RAP case.}
   \end{figure}

On Fig. \ref{fig:VLT_Fig4}, we extend the piston petal-level error study to a large range of aberration amplitudes, from $10$ mrad RMS to $5$ rad RMS. The red plot corresponds to the SPHERE coronagraph case and the green one to the RAP case. In the SPHERE coronagraph case, the contrast degrades from $1.2 \times 10^{-6}$ to $1.6 \times 10^{-5}$ (factor of $13$) while in the RAP case the contrast stabilizes around $4.1 \times 10^{-7}$ (degradation of a factor of $1.7$), which indicates an increased robustness compared to the SPHERE coronagraph.

%-------------
%-------------

%%%%%%%%%%%%%%%%%%%%%%%%%%%%%%%%%%%%%%%%%%%%%%%%%%%%%%%%%%%%%%%%%
\subsection{Other application}
\label{sec:Other application}
%%%%%%%%%%%%%%%%%%%%%%%%%%%%%%%%%%%%%%%%%%%%%%%%%%%%%%%%%%%%%%%%%

\begin{figure}%[h]
   \begin{center}
   \begin{tabular}{c}
   \includegraphics[width=12cm]{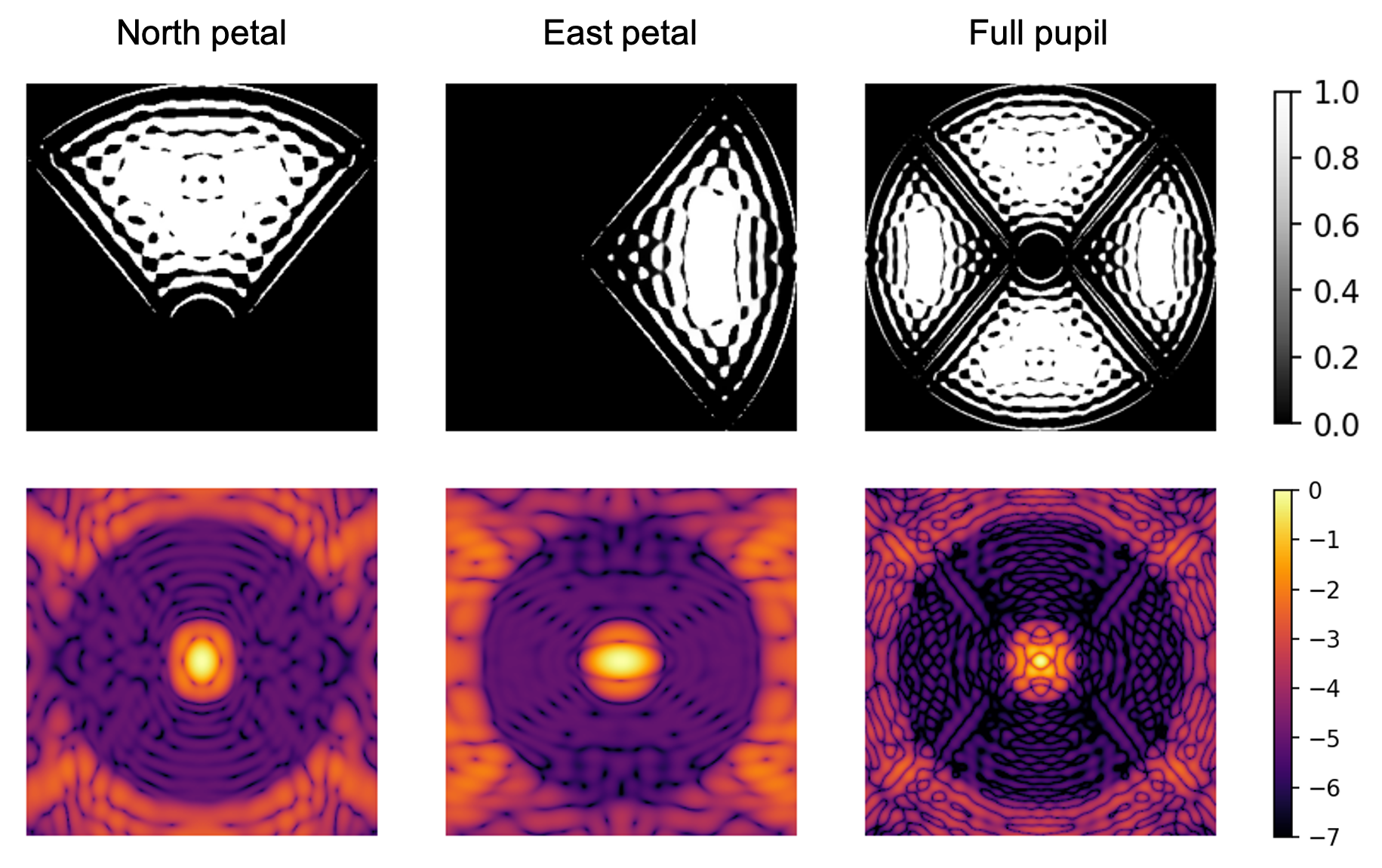}
   \end{tabular}
   \end{center}
   \caption[VLT_Fig1bis] 
   { \label{fig:VLT_Fig1bis} SP design from redundant apodized petals: (top) apodized petals (North and East) and recombined RAP, (bottom) contrast maps in logarithmic scale generated by the apodizers of the first line.}
   \end{figure}

We design a RAP that can access a maximum contrast of $1.0 \times 10^{-5}$ between $6 \lambda/D$ and $20 \lambda/D$. The two petal apodizers and the RAP after recombination are shown on Fig. \ref{fig:VLT_Fig1bis}. Despite releasing the deep contrast constraint compared to the previous case (this RAP dark zone has an average contrast of $1.7 \times 10^{-6}$), the gain in IWA impacts the planetary throughput ($28\%$) and transmission ($51\%$).

%-------------
%-------------

%\color{red}
%- on SPHERE, mean contrast deteriorated from $1.6 \times 10^{-6}$ to $3.8 \times 10^{-5}$ (factor $24$) versus on RAP from $1.7 \times 10^{-6}$ to $3.2 \times 10^{-6}$ (factor lower than $2$). Fig. \ref{fig:VLT_Fig2bis}
%\color{black}

\begin{figure}%[h]
   \begin{center}
   \begin{tabular}{c}
   \includegraphics[width=12cm]{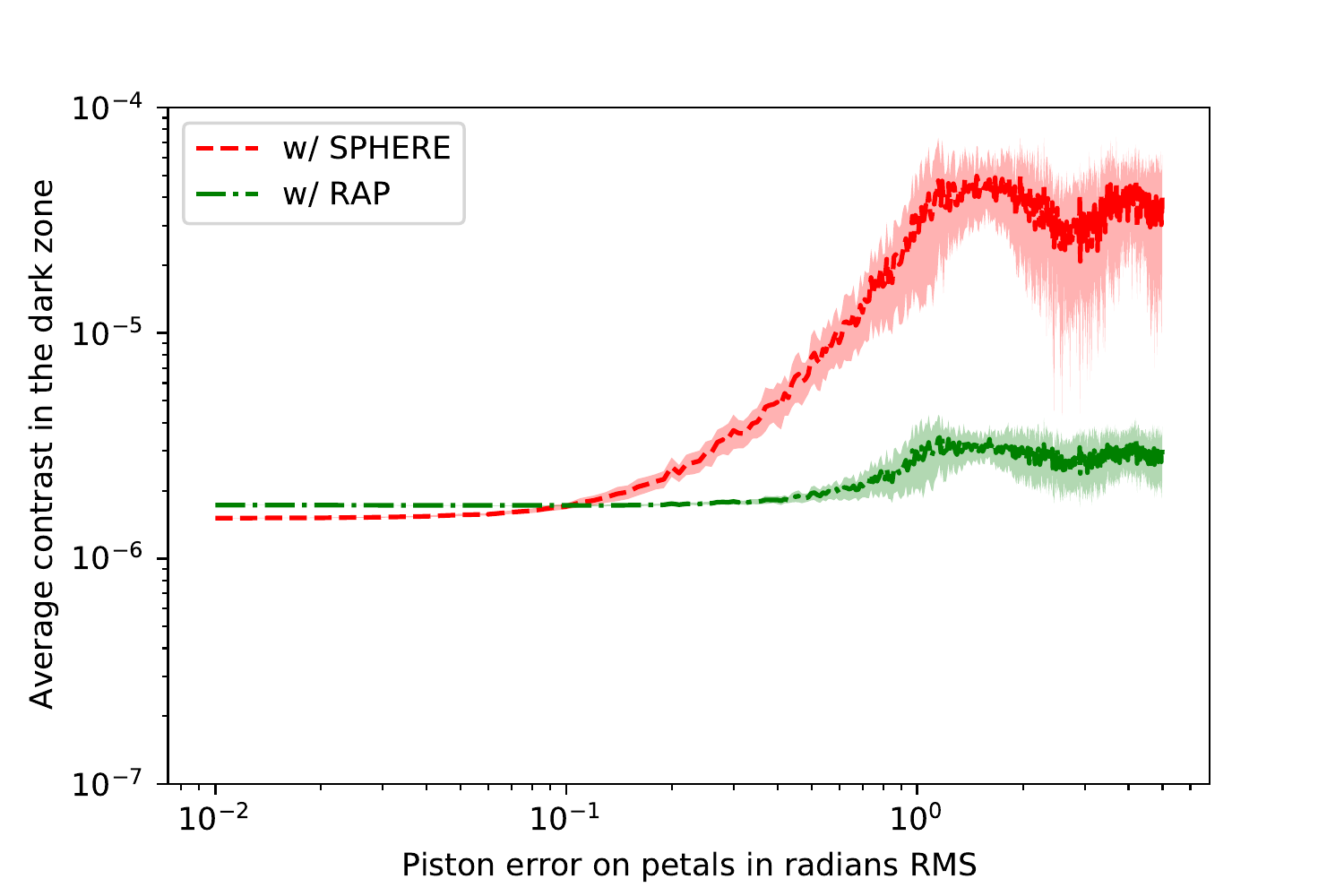}
   \end{tabular}
   \end{center}
   \caption[VLT_Fig4bis] 
   { \label{fig:VLT_Fig4bis} Evolution of the average contrast in the dark zone with the amplitude of the petal-level piston errors: for each amplitude, $25$ random phase maps are propagated through the optical system and the mean value of the $25$ resulting dark zone contrasts is computed (dark lines). The light color areas are delimited by the standard deviations of the $25$ dark zone contrasts. The red plots correspond to the SPHERE coronagraph case and the green ones to the RAP case.}
   \end{figure}

In Fig. \ref{fig:VLT_Fig4bis}, we plot the evolution of the mean contrast in the dark zone as a function of the piston petal-level aberration amplitude, with both the SPHERE coronagraph (red) and this new RAP design (green). We obtain similar results than the ones of section \ref{sec:Error budget2}: for aberrations from $10$ mrad RMS to $5$ rad RMS, the SPHERE coronagraph average contrast deteriorates from $1.5\times 10^{-6}$ to $3.5 \times 10^{-5}$ (factor of $23$) and the RAP average contrast from $1.7\times 10^{-6}$ to $2.9 \times 10^{-6}$ (factor of $1.7$). Once again, the gain of a redundant SP is significant and the impact of low-wind effect on the coronagraphic performance remains low despite aberrations up to several radians RMS.

%-------------
%-------------

%%%%%%%%%%%%%%%%%%%%%%%%%%%%%%%%%%%%%%%%%%%%%%%%%%%%%%%%%%%%%%%%%
%%%%%%%%%%%%%%%%%%%%%%%%%%%%%%%%%%%%%%%%%%%%%%%%%%%%%%%%%%%%%%%%%
%%%%%%%%%%%%%%%%%%%%%%%%%%%%%%%%%%%%%%%%%%%%%%%%%%%%%%%%%%%%%%%%%
\section{Conclusions and perspectives}
\label{sec:Conclusions}
%%%%%%%%%%%%%%%%%%%%%%%%%%%%%%%%%%%%%%%%%%%%%%%%%%%%%%%%%%%%%%%%%
%%%%%%%%%%%%%%%%%%%%%%%%%%%%%%%%%%%%%%%%%%%%%%%%%%%%%%%%%%%%%%%%%
%%%%%%%%%%%%%%%%%%%%%%%%%%%%%%%%%%%%%%%%%%%%%%%%%%%%%%%%%%%%%%%%%

This proceeding follows the development of redundant apodizers first introduced in Leboulleux et al. 2022\cite{Leboulleux2022} for segmentation-due errors in a GMT-like case and then applied to island effects (low-wind effect and post-adaptive optics system petaling) on a ELT-like telescope\cite{Leboulleux2022a}. In this proceeding, we propose other applications to redundant apodizers: first, the segmentation-due error robustness is applied on a telescope with more segments (18 hexagonal segments) for a localized dark zone compatible with spectroscopy for instance with an Integral Field Unit; and secondly, the robustness to island effects is applied on the VLT configuration, for two large circular dark zones.  

Designing a coronagraph implies a trade-off between different parameters including the IWA, the contrast, the throughput, and the robustness to aberrations, more often evaluated after design. The RAP concept allows to optimize the coronagraph design to make it directly robust to segment- and petal-level errors, but with a loss on other parameters: no optimization at angular separations smaller than the segment or petal diffraction limit and a decrease of throughput for a given contrast compared to full-pupil apodizers. As a compromise between performance and stability, a second design step, with an optimization on the full pupil, could be added to the segment- or petal-level apodization, to improve the contrast and/or access smaller angular separations with no robustness at these separations though.

Despite this trade-off, RAPs allow a coronagraph to perform in non optimal conditions, meaning in presence of segment phasing errors, low-wind effect, and post-adaptive optics petaling. These aberrations are already limiting some current high-contrast imagers (VLT/SPHERE, Subaru/SCExAO) and will limit the ones under development (ELT, LUVOIR...). The RAP concept is a passive solution compatible with most of these telescopes to access the coronagraph deep contrast in presence of such errors. 

\acknowledgments 
This project is funded by the European Research Council (ERC) under the European Union's Horizon 2020 research and innovation programme (grant agreement n°866001).\\
This work has also been partially supported by the LabEx FOCUS ANR-11-LABX-0013.

\bibliography{bib}
\bibliographystyle{spiebib}

\end{document}